\documentclass{article}
\usepackage{amsmath,amsthm,amsfonts,graphicx,amssymb}
\usepackage{url,xspace}
\usepackage{algorithm,%
algpseudocode%
}
\usepackage[T1]{fontenc}
\usepackage[utf8]{inputenc}

\renewcommand{\algorithmicrequire}{{\textbf{Input:}}}
\renewcommand{\algorithmicensure}{{\textbf{Output:}}}
\usepackage{rotating}

\newcommand{\U}[1]{{\ensuremath{\bf{U_#1}}}}
\newcommand{\GO}[1]{\ensuremath{\mathcal{O}\left(#1\right)}\xspace}
\newcommand{\dhss}{\cite{Douglas:1994:gemmw}\xspace}
\newcommand{\hljjtt}{\cite{Huss-Lederman:1996:mai}\xspace}
\newcommand{\ip}{\texttt{IP}\xspace}
\newcommand{\ipl}{\texttt{OvL}\xspace}
\newcommand{\ipr}{\texttt{OvR}\xspace}
\newcommand{\acco}{\texttt{AcLR}\xspace}

\newcommand{\accr}{\texttt{AccR}\xspace}
\newcommand{\acc}{\texttt{Acc}\xspace}
\newcommand{\accrb}{\texttt{AcR}\xspace}
\newcommand{\IP}{\texttt{IP}\xspace}
\renewcommand{\geq}{\geqslant}
\renewcommand{\leq}{\leqslant}
\newtheorem{thm}{Theorem}
\newtheorem{lem}{Lemma}

\newtheorem{prop}{Proposition}

\usepackage{paralist,array,enumerate,multirow,float}
\def\ie{\mbox{{i.e.}}}         
\urldef\jgdemail\url{\{Brice.Boyer,Jean-Guillaume.Dumas\}@imag.fr}
\urldef\wzemail\url{w2zhou@uwaterloo.ca}
\urldef\cpemail\url{Clement.Pernet@imag.fr}

\begin{document}
\title{Memory efficient scheduling of Strassen-Winograd's matrix
  multiplication algorithm\footnote{\copyright ACM, 2009. This is the
    author's version of the work. It is posted here by permission of
    ACM for your personal use. Not for redistribution. The definitive
    version was published in ISSAC 2009.}}
\author{Brice Boyer\thanks{Laboratoire J. Kuntzmann, Universit\'e de
  Grenoble. 51, rue des Math\'ematiques, umr CNRS 5224, bp 53X, F38041
  Grenoble, France, \jgdemail}
\and Jean-Guillaume Dumas\footnotemark[1]
\and Cl\'ement Pernet\thanks{Laboratoire LIG, Universit\'e de
  Grenoble. umr CNRS, F38330 Montbonnot, France. \cpemail}
\and Wei Zhou\thanks{School of Computer Science, University of
  Waterloo, Waterloo, ON, N2B 3G1, Canada. \wzemail}
}
\maketitle

\begin{abstract}
We propose several new schedules for Strassen-Winograd's matrix multiplication algorithm, they reduce the extra memory allocation requirements by three different
means: by introducing a few
pre-additions, by overwriting the input matrices, or by using a
first recursive
level of classical multiplication.
In particular, we show two fully in-place schedules: one having the
same number of
operations, if the input matrices can be overwritten; the other one,
slightly increasing
the constant of the leading term of the complexity, if the input matrices are
read-only. 
Many of these schedules have been found by an implementation of an exhaustive
search algorithm based on a pebble game.
\end{abstract}
%


\section{Introduction}
Strassen's algorithm~\cite{Strassen:1969:GENO} was the
first sub-cubic algorithm for matrix multiplication.
Its improvement by Winograd~\cite{Winograd:1971:tbtmm}
led to a highly practical algorithm.
The best asymptotic complexity for this computation has been
successively improved since then, down to $\GO{n^{2.376}}$
in \cite{Coppersmith:1990:MMAP} (see
\cite{Bini:1994:PMC-FA,burgisser:1997} for a review), but
Strassen-Winograd's
still remains one of the most practicable.
Former studies on how to turn this algorithm into practice can be found
in~\cite{bailey:603,Huss-Lederman:1996:ISA, Huss-Lederman:1996:mai,
  Douglas:1994:gemmw}
and references therein for numerical computation and in
\cite{Pernet:2001:Winograd, Dumas:2002:FFLAS}
for computations over a finite field.\\
In this paper, we propose new schedules of the algorithm, that
reduce the
extra memory allocation, by three different means:
by introducing a few pre-additions, by overwriting the input matrices, or by using a first recursive level of classical multiplication.
These schedules can prove useful for instance for memory
efficient computations of the rank,
determinant, nullspace basis, system resolution, matrix inversion...
Indeed, the matrix multiplication based LQUP factorization of~\cite{Ibarra:1982:LSP} can be
computed with no other temporary allocations than the ones involved in
its block matrix
multiplications~\cite{JeannerodPernet:2007:LQUP}. Therefore the
improvements on
the memory requirements of the matrix multiplication, used together
for instance with cache optimization strategies~\cite{Bader:2006:comm}, will directly
improve these higher level computations.

We only consider here the computational complexity and space
complexity, counting the number of arithmetic operations and memory
allocations.
The focus here is neither on stability issues, nor really on speed
improvements. We rather study potential memory space savings.
Further studies have thus to be made to assess for some gains for
in-core computations or to use these schedules
for numerical computations.
They are nonetheless already useful for exact computations, for instance on
integer/rational or finite field
applications~\cite{Dumas:2004:FFPACK,Laderman:1992:PAA}. 

The remainder of this paper is organized as follows: we review Strassen-Winograd's algorithm and existing memory schedules
in sections \ref{sec:algo} and \ref{ssec:constinput}. We then present in
section \ref{sec:pebble} the dynamic program we used to search for
schedules. This allows us to give several schedules overwriting their inputs in
section \ref{sec:overwrite}, and then a new schedule for $C \leftarrow
A B +C$ using only two extra temporaries in section \ref{sec:hyb}, all
of them preserving the leading term of the arithmetic complexity. 
Finally, in section \ref{sec:mix}, we present a generic way of transforming non in-place matrix
multiplication algorithms into in-place ones (\ie{} without any extra
temporary space), with a small constant
factor overhead. Then we recapitulate in table \ref{tab:resume} the
different available schedules and give their respective features.
%
\section{Strassen-Winograd Algorithm}\label{sec:algo}
We first review Strassen-Winograd's algorithm, and setup the notations that
will be used throughout the paper.\\
Let $m,n$ and $k$ be powers of $2$.
Let $A$ and $B$ be two matrices of dimension $m\times k$ and $k\times n$ and
let $C=A\times B$.
Consider the natural block decomposition:
$$
\begin{bmatrix}
  C_{11} & C_{12}\\
  C_{21} & C_{22}\\
\end{bmatrix}
=
\begin{bmatrix}
  A_{11} & A_{12}\\
  A_{21} & A_{22}\\
\end{bmatrix}
\begin{bmatrix}
  B_{11} & B_{12}\\
  B_{21} & B_{22}\\
\end{bmatrix},
$$
where $A_{11}$ and $B_{11}$ respectively have dimensions $m/2 \times k/2$ and $k/2
\times n/2$.
Winograd's algorithm computes the $m\times n$ matrix $C=A\times B$ with the
following 22 block operations:\\
\begin{inparaenum}[$\bullet$\xspace]
\item 8 additions:
$$\begin{array}{lll}
  S_1  \leftarrow A_{21} + A_{22}	& S_2  \leftarrow S_1 - A_{11}
  & S_3  \leftarrow A_{11} - A_{21} \\
T_1 \leftarrow B_{12} - B_{11} & T_2 \leftarrow B_{22} - T_1 & T_3 \leftarrow B_{22} - B_{12} \\
  S_4  \leftarrow A_{12} - S_2		&				& T_4 \leftarrow T_2 - B_{21}	 \\
\end{array}$$
%
\item 7 recursive multiplications:
$$\begin{array}{lll}
  P_1  \leftarrow A_{11} \times B_{11}  &  P_2  \leftarrow A_{12}
  \times B_{21}  \\
P_3  \leftarrow S_4 \times B_{22}     &P_4  \leftarrow A_{22} \times
T_4	\\
P_5 \leftarrow S_1 \times T_1 & P_6\leftarrow S_2 \times T_2 & P_7 \leftarrow S_3 \times T_3 \\
 \end{array}$$%
%
\item 7 final additions:
$$\begin{array}{lll}
  U_1  \leftarrow P_1 + P_2 & U_2  \leftarrow P_1 + P_6 \\
U_3  \leftarrow U_2 + P_7 &U_4  \leftarrow U_2 + P_5\\
U_5 \leftarrow U_4 + P_3 & U_6 \leftarrow U_3 - P_4& U_7 \leftarrow U_3 + P_5  \\
\end{array}$$
%
\item The result is the matrix:
{ $ C = \left[ \begin{array}{ll} U_1 & U_5 \\ U_6 & U_7 \end{array} \right]$}.
\end{inparaenum}

\begin{figure}[htb]
\begin{center}
\includegraphics[width=\columnwidth]{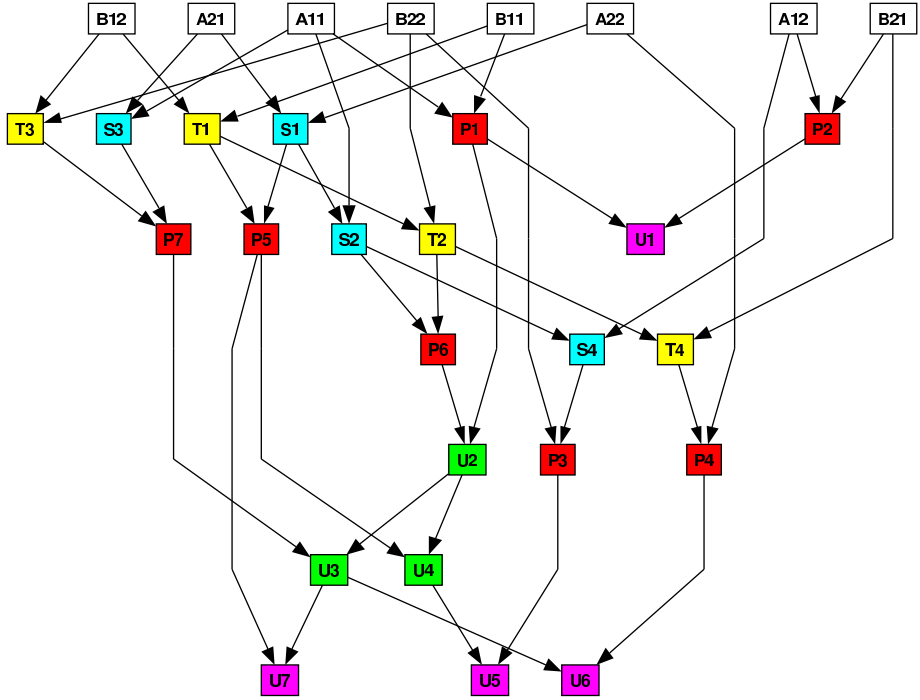}
\caption{Winograd's task dependency graph}
\label{fig:winotask}
\end{center}
\end{figure}

Figure~\ref{fig:winotask} illustrates the dependencies between these tasks.
%
%
\section{Existing memory placements}%
\label{ssec:constinput}
Unlike the classic multiplication algorithm,
Winograd's algorithm requires some extra temporary memory
allocations to perform its 22 block operations.
%
\subsection{Standard product}\label{sub:exist:standard}
We first consider the basic operation $C\leftarrow A\times B$. The best known
schedule for this case was given by
\cite{Douglas:1994:gemmw}. We reproduce a
similar schedule in table~\ref{tab:schedule:AB}.
%
\begin{table}[htb]
	\newcolumntype{s}{>{\!\!\scriptsize}l<{\!\!\!\!}}
	\newcolumntype{t}{>{\!\!\!\!\scriptsize}r<{\!\!}}
	\small
	\begin{center}
		\begin{tabular}{|slt|slt|}
			\hline
			\# & operation & loc. & \# & operation & loc.\\
			\hline
			1  & $S_3 = A_{11} - A_{21}$	& $X$		& 12 & $P_1 = A_{11} B_{11}$	& $X$ \\
			2  & $T_3 = B_{22} - B_{12}$	& $Y$		& 13 & $U_2 = P_1 + P_6$		& $C_{12}$\\
			3  & $P_7 = S_3 T_3$			& $C_{21}$	& 14 & $U_3 = U_2 + P_7$		& $C_{21}$ \\
			4  & $S_1 = A_{21} + A_{22}$	& $X$		& 15 & $U_4 = U_2 + P_5$		& $C_{12}$ \\
			5  & $T_1 = B_{12} - B_{11}$	& $Y$		& 16 & $\U7 = U_3 + P_5$		& $C_{22}$ \\
			6  & $P_5 = S_1 T_1$			& $C_{22}$	& 17 & $\U5 = U_4 + P_3$		& $C_{12}$ \\
			7  & $S_2 = S_1 - A_{11}$		& $X$		& 18 & $T_4 = T_2 - B_{21}$ 	& $Y$\\
			8  & $T_2 = B_{22} - T_1$		& $Y$		& 19 & $P_4 = A_{22} T_4$		& $C_{11}$\\
			9  & $P_6 = S_2 T_2$			& $C_{12}$	& 20 & $\U6 = U_3 - P_4$		& $C_{21}$ \\
			10 & $S_4 = A_{12} - S_2$		& $X$		& 21 & $P_2 = A_{12} B_{21}$	& $C_{11}$  \\
			11 & $P_3 = S_4 B_{22}$			& $C_{11}$  & 22 & $\U1 = P_1 + P_2$		& $C_{11}$\\
			\hline
		\end{tabular}
		\caption{Winograd's algorithm for operation $C\leftarrow A\times B$, with two temporaries}
		\label{tab:schedule:AB}
	\end{center}
\end{table}
%
It requires two temporary blocks $X$ and $Y$ whose dimensions are respectively equal to
$m/2\times \max(k/2, n/2)$ and $k/2 \times n/2$.
Thus the extra memory used is:
$$E_{\ref{tab:schedule:AB}}(m,k,n)= \frac{m}{2} \max{\left(\frac{k}{2}, \frac{n}{2}\right)} + \frac{k}{2} \frac{n}{2}+ E_{\ref{tab:schedule:AB}}\left(\frac{m}{2}, \frac{k}{2}, \frac{n}{2}\right).$$
Summing these temporary allocations over every recursive levels leads to a total amount of memory, where for brevity $M = \min{\{m,k,n\}}$:
\begin{align}
	E_{\ref{tab:schedule:AB}}(m,k,n) & = \sum_{i=1}^{\log_2(M)} \frac{1}{4^i}  \left(m \max{(k,n)} + k n\right) \label{Eq:schedule:AB} \\
	& = \frac{1}{3} \left( 1 - \frac{1}{M^2}\right) \left(m \max{(k,n)} + k n\right) \notag \\
	& < \frac{1}{3}  \left(m \max{(k,n)} + k n\right). \notag
\end{align}
We can prove in the same manner the following lemma:
\begin{lem}\label{lem:sum}
	Let $m$, $k$ and $n$  be powers of two, $g(x,y,z)$ be homogeneous, $M =  \min{\{m,k,n\}}$ and $f(m,k,n)$ be a function such that 
	$$f(m,k,n) = \begin{cases} 
		g\left(\frac{m}{2}, \frac{k}{2}, \frac{n}{2}\right) + f\left(\frac{m}{2}, \frac{k}{2}, \frac{n}{2}\right)	& \text{if\ } m\text{,\,}n\text{\,and\,}k > 1\\
		0																& \text{{otherwise.}}
	\end{cases}$$ 
Then $f\left(m,k,n\right) = \frac{1}{3}  \left( 1 - \frac{1}{M^2}\right) g(m,k,n)< \frac{1}{3} g(m,k,n)$.
\end{lem}
In the remainder of the paper, we use $E_i$ to denote the amount of extra memory used in table number~$i$. The amount of extra memory we consider is always the sum up to the last recursion level.

Finally, assuming $m=n=k$  gives a total extra memory requirement of
$E_{\ref{tab:schedule:AB}}(n,n,n) < 2/3 n^2.$
\subsection{Product with accumulation}
For the more general operation $C \leftarrow \alpha A\times B +\beta C$,
a first na\"ive method would compute the product $\alpha A\times B$ using
the  scheduling of table~\ref{tab:schedule:AB}, into a temporary matrix $C'$
and finally compute  $C\leftarrow C'+\beta C$. It would require $(1+2/3)n^2$ extra
memory allocations in the square case.\\
Now the schedule of table~\ref{tab:schedule:ABC} due to
~\cite[fig. 6]{Huss-Lederman:1996:mai} only  requires 3 temporary blocks for the
same number of operations ($7$ multiplications and $4+15$ additions).
%
\begin{table}[htb]
	\small
	\newcolumntype{s}{>{\!\!\scriptsize}l<{\!\!\!\!}}
	\newcolumntype{t}{>{\!\!\!\!\scriptsize}r<{\!\!}}
	\begin{center}
		\begin{tabular}{|slt|slt|}
			\hline
			\# & operation & loc. & \# & operation & loc.  \\
			\hline
			1 & $S_1 = A_{21} + A_{22}$			& $X$		& 12 & $S_4 = A_{12} - S_2$						& $X$ \\
			2 & $T_1 = B_{12} - B_{11}$			& $Y$		& 13 & $T_4 = T_2 - B_{21}$						& $Y$ \\
			3 & $P_5 = \alpha S_1 T_1 $			& $Z$		& 14 & $C_{12} =  \alpha S_4 B_{22} + C_{12}$	& $C_{12}$ \\
			4 & $ C_{22} = P_5 + \beta C_{22}$	& $C_{22}$	& 15 & $\U5 = U_2 + C_{12} $					& $C_{12}$\\
			5 & $ C_{12} = P_5 + \beta C_{12}$	& $C_{12}$	& 16 & $P_4 = \alpha A_{22}T_4-\beta C_{21}$	& $C_{21}$\\
			6 & $S_2 = S_1 - A_{11}$			& $X$		& 17 & $S_3 = A_{11} - A_{21}$					& $X$   \\
			7 & $T_2= B_{22} - T_1$				& $Y$		& 18 & $T_3 = B_{22} - B_{12}$					& $Y$   \\
			8 & $P_1 = \alpha A_{11} B_{11}$	& $Z$		& 19 & $U_3 = \alpha S_3 T_3+U_2$				& $Z$\\
			9 & $C_{11} = P_1 + \beta C_{11}$	& $C_{11}$	& 20 & $\U7 = U_3 + C_{22}$						& $C_{22}$ \\
			10& $U_2 = \alpha S_2 T_2 + P_1$	& $Z$		& 21 & $\U6 = U_3 - C_{21}$						& $C_{21}$ \\
			11&$\U1 =\alpha A_{12}B_{21}+C_{11}$& $C_{11}$	& 22 & &\\
			\hline
		\end{tabular}
		\caption{Schedule for operation $C\leftarrow \alpha A\times B+\beta C$ with 3 temporaries}
		\label{tab:schedule:ABC}
	\end{center}
\end{table}
The required three temporary blocks $X, Y, Z$ have dimensions $m/2\times
k/2$, $k/2 \times n/2$ and $m/2 \times n/2$. 
Since the two temporary blocks in schedule~\ref{tab:schedule:AB} 
are smaller than the three ones here, we have $E_2 \geq E_1$.
Hence, using lemma~\ref{lem:sum}, we get
\begin{equation}\label{Eq:schedule:ABC}
	E_{\ref{tab:schedule:ABC}}\left(m,k,n\right) = \frac{1}{3} \left( 1 - \frac{1}{M^2}\right) \left(mk+kn+mn \right).
\end{equation}
With $m=n=k$, this gives $E_{\ref{tab:schedule:ABC}}(n,n,n) < n^2.$\\
%
We propose in table~\ref{tab:ABC:2tmp} a new schedule for the same operation
$\alpha A\times B + \beta C$ only requiring two temporary blocks.\\
Our new schedule is more efficient if some inner calls overwrite their
temporary input matrices. We now present some overwriting schedules
and the dynamic program we used to find them.
\section{Exhaustive search algorithm}\label{sec:pebble}
We used a brute force search algorithm\footnote{The code is available
  at \url{http://ljk.imag.fr/CASYS/LOGICIELS/Galet}.} to get some of the new
schedules that will be presented in the following sections.
It is very similar to the pebble game of
Huss-Lederman et al.\ \cite{Huss-Lederman:1996:mai}.\\
A sequence of computations is represented as a directed graph, just like figure~\ref{fig:winotask} is built from Winograd's algorithm.\\
A node represents a program variable. The nodes can be classified as initials (when they correspond to inputs), temporaries
(for intermediate computations) or finals (results or nodes that we want to
keep, such as ready-only inputs).\\
The edges represent the operations; they point from the operands to the result.\\
A pebble represents an allocated memory. We can put pebbles on any
nodes, move or remove them according to a set of simple rules shown below.\\
When a pebble arrives to a node, the computation at the associated
variable starts, and can be ``partially'' or ``fully''
executed. If not specified, it is assumed that the computation is
fully executed.\\
Edges can be removed, when the corresponding operation has been
computed.\\
The last two points are especially useful for accumulation
operations: for example, it is possible to try schedule the multiplication
separately from the addition in an otherwise recursive $AB+C$ call;
the edges involved in the multiplication operation would then be removed first and the accumulated part later.
They are also useful if we do not want to fix the way some additions are
performed: if $U_3 = P_1 + P_6  + P_7$ the associativity allows
different ways of computing the sum and we let the program explore
these possibilities.
At the beginning of the exploration, each initial node has a pebble
and we may have a few extra available pebbles. 
The program then tries to apply the following rules, in order, on each
node. The program stops when every final node has a pebble or when
no further moves of pebbles are possible:

\begin{inparaenum}[{\ $\bullet$\xspace\emph{Rule}} 1.]
\item[{\ $\bullet$\xspace\emph{Rule}} 0.] \emph{Computing a result/removing edges.} If a node
has a pebble and parents with pebbles, then the operation can be
performed and the corresponding edges removed. The node is
then at least partially computed.

\item \emph{Freeing some memory/removing a pebble.} If a node is
isolated and not final, its pebble is freed. This means that
we can reclaim the memory here because this node has been fully
computed (no edge pointing to it) and is no longer in use as an
operand (no edge initiating from~it).

\item \emph{Computing in place/moving a pebble.} If a node $P$ has
a full pebble and a single empty child node $S$ and if other
parents of $S$ have pebbles on them, then the pebble on $P$ may
be transferred  to $S$ (corresponding edges are removed). This
means an operation has been made in place in the parent $P$'s pebble.

\item \emph{Using more memory/adding a pebble.} If parents of an empty node $N$ have
pebbles and a free pebble is available, then this pebble can be assigned to $N$ and the
corresponding edges are removed. This means that the operation is
computed in a new memory location.

\item \emph{Copying some memory/duplicating a pebble.} 
A computed node having a pebble can be duplicated. The edges pointed
to or from the original node are then rearranged between them.
This means that a temporary result has been copied into some free
place to allow more flexibility.
\end{inparaenum}
\section{Overwriting input matrices}
\label{sec:overwrite}
We now relax some constraints on the previous problem: the input matrices $A$
and $B$ can be overwritten, as proposed
by~\cite{Kreczmar:1976:Strassen}.
For the sake of simplicity, 
we first give schedules only working for square matrices (i.e. $m=n=k$
and any memory location is supposed to be able to receive any result
of any size).
We nevertheless give the memory requirements of
each schedule as a function of $m$; $k$ and $n$.
Therefore it is easier in the last part of this section to adapt the proposed schedules partially
for the general case. 
In the tables, the notation $A_{ij} B_{ij}$ (resp. $A_{ij} B_{ij} +
C_{ij})$ denotes the use of the algorithm from
table~\ref{tab:schedule:AB} (resp. table~\ref{tab:schedule:ABC}) as a
subroutine. Otherwise we use the notation $Alg(A_{ij} B_{ij})$ to
denote a recursive call or the use of one of our new schedules as a
subroutine.
%
\subsection{Standard product}
We propose in table~\ref{tab:AB:inplace} a new schedule
that computes the
product $C\leftarrow A\times B$ without any temporary memory
allocation. The idea here is to find an ordering where the recursive
calls can be made also in place such that the operands of a
multiplication are no longer in use after the multiplication has completed because they are overwritten.
An exhaustive search showed that no schedule exists overwriting less
than four sub-blocks.
%
%
\begin{table}[htb]
	\newcolumntype{s}{>{\!\!\scriptsize}l<{\!\!\!\!}}
	\newcolumntype{t}{>{\!\!\!\!\scriptsize}r<{\!\!}}
	\small
	\begin{center}
		\begin{tabular}{|slt|slt|}
			\hline
			\# & operation & loc. & \# & operation & loc.  \\
			\hline
			1  & $S_3 = A_{11} - A_{21}$	& $C_{11}$ & 12 & $S_4 = A_{12} - S_2$			& $A_{22}$ \\
			2  & $S_1 = A_{21} + A_{22}$	& $A_{21}$ & 13 & $P_6 = \IP(S_2 T_2)$			& $C_{22}$ \\
			3  & $T_1 = B_{12} - B_{11}$	& $C_{22}$ & 14 & $U_2 = P_1 + P_6$				& $C_{22}$ \\
			4  & $T_3 = B_{22} - B_{12}$	& $B_{12}$ & 15 & $P_2 = \IP(A_{12} B_{21})$	& $C_{12}$ \\
			5  & $P_7 = \IP(S_3 T_3$)		& $C_{21}$ & 16 & $\U1 = P_1 + P_2$				& \bf $C_{11}$ \\
			6  & $S_2 = S_1 - A_{11}$		& $C_{12}$ & 17 & $U_4 = U_2 + P_5$				& $C_{12}$ \\ 
			7  & $P_1 = \IP(A_{11} B_{11})$	& $C_{11}$ & 18 & $U_3 = U_2 + P_7$				& $C_{22}$ \\
			8  & $T_2 = B_{22} - T_1$		& $B_{11}$ & 19 & $\U6 = U_3 - P_4$				& \bf $C_{21}$ \\
			9  & $P_5 = \IP(S_1 T_1)$		& $A_{11}$ & 20 & $\U7 = U_3 + P_5$				& \bf $C_{22}$ \\
			10 & $T_4 = T_2 - B_{21}$		& $C_{22}$ & 21 & $P_3 = \IP(S_4 B_{22})$		& $A_{12}$ \\
			11 & $P_4 = \IP(A_{22} T_4)$	& $A_{21}$ & 22 & $\U5 = U_4 + P_3$				& \bf $C_{12}$ \\
			\hline
		\end{tabular}
		\caption{\IP schedule for operation $C\leftarrow A\times B$ in place}
		\label{tab:AB:inplace}
	\end{center}
\end{table}
Note that this schedule uses only two blocks of $B$ and the whole of $A$
but overwrites all of $A$ and $B$.
For instance the recursive computation of $P_2$ requires overwriting parts of $A_{12}$ and $B_{21}$ too. 
%
Using another schedule as well as back-ups of overwritten parts into some available memory 
%
%
%
In the following, we will denote by \ip for {\texttt InPlace}, either one of
these two schedules.\\
We present in tables~\ref{tab:AB:ipleft} and
\ref{tab:AB:ipright} two new schedules overwriting only one of
the two input matrices, but requiring an extra temporary space.
These two schedules are denoted \ipl and \ipr.
The exhaustive search also showed that no schedule
exists overwriting only one of $A$ and $B$ and using no extra temporary.
%
\begin{table}[htb]
	\small
	\newcolumntype{s}{>{\!\!\scriptsize}l<{\!\!\!\!}}
	\newcolumntype{t}{>{\!\!\!\!\scriptsize}r<{\!\!}}
	\begin{center}
		\begin{tabular}{|slt|slt|}
			\hline
			\# & operation & loc. & \# & operation & loc.  \\
			\hline
			1  & $S_3 = A_{11} - A_{21}$		& $C_{22}$	& 12 & $P_6 = \ipl(S_2 T_2)$			& $C_{21}$		\\
			2  & $S_1 = A_{21} + A_{22}$		& $A_{21}$	& 13 & $T_4 = T_2 - B_{21}$				& $A_{11}$		\\
			3  & $S_2 = S_1 - A_{11}$			& $C_{12}$	& 14 & $U_2 = P_1 + P_6$				& $C_{21}$		\\
			4  & $T_1 = B_{12} - B_{11}$		& $C_{21}$	& 15 & $U_4 = U_2 + P_5$				& $C_{12}$		\\
			5  & $P_1 = \ipl(A_{11} B_{11})$	& $C_{11}$	& 16 & $U_3 = U_2 + P_7$				& $C_{21}$		\\
			6  & $T_3 = B_{22} - B_{12}$		& $A_{11}$	& 17 & $\U7 = U_3 + P_5$				& \bf $C_{22}$	\\
			7  & $P_7 = \ip(S_3 T_3)$			& $X$		& 18 & $\U5 = U_4 + P_3$				& \bf $C_{12}$	\\	 
			8  & $T_2 = B_{22} - T_1$			& $A_{11}$	& 19 & $P_2 = \ipl(A_{12} B_{21})$		& $X$			\\
			9  & $P_5 = \ip(S_1 T_1)$			& $C_{22}$	& 20 & $\U1 = P_1 + P_2$				& \bf $C_{11}$	\\ 
			10 & $S_4 = A_{12} - S_2$			& $C_{21}$	& 21 & $P_4 = \IP(A_{22} T_4)$			& $A_{21}$		\\ 
			11 & $P_3 = \ipl(S_4 B_{22})$		& $A_{21}$	& 22 & $\U6 = U_3 - P_4$				& \bf $C_{21}$  \\
			\hline
		\end{tabular}
		\caption{\ipl schedule for operation $C\leftarrow A\times B$ using strictly two
		blocks of $A$ and one temporary}
		\label{tab:AB:ipleft}
	\end{center}
\end{table}
%
%
%
\begin{table}[htb]
	\small
	\newcolumntype{s}{>{\!\!\scriptsize}l<{\!\!\!\!}}
	\newcolumntype{t}{>{\!\!\!\!\scriptsize}r<{\!\!}}
	\begin{center}
		\begin{tabular}{|slt|slt|}
			\hline
			\# & operation & loc. & \# & operation & loc.  \\
			\hline
			1  & $S_3 = A_{11} - A_{21}$		& $C_{22}$	& 12 & $P_4 = \ipr(A_{22} T_4)$		& $B_{12}$ \\	
			2  & $S_1 = A_{21} + A_{22}$		& $C_{21}$	& 13 & $S_4 = A_{12} - S_2$				& $B_{11}$ \\
			3  & $T_1 = B_{12} - B_{11}$		& $C_{12}$	& 14 & $U_2 = P_1 + P_6$				& $C_{21}$ \\
			4  & $P_1 = \ipr(A_{11} B_{11})$	& $C_{11}$	& 15 & $U_4 = U_2 + P_5$				& $C_{12}$  \\
			5  & $S_2 = S_1 - A_{11}$			& $B_{11}$	& 16 & $U_3 = U_2 + P_7$				& $C_{21}$ \\
			6  & $T_3 = B_{22} - B_{12}$		& $B_{12}$	& 17 & $\U7 = U_3 + P_5$				& \bf $C_{22}$ \\
			7  & $P_7 = \ip(S_3 T_3)$			& $X$		& 18 & $\U6 = U_3 - P_4$				& \bf $C_{21}$\\
			8  & $T_2 = B_{22} - T_1$			& $B_{12}$	& 19 & $P_3 = \IP(S_4 B_{22})$			& $B_{12}$ \\
			9  & $P_5 = \ip(S_1 T_1)$			& $C_{22}$	& 20 & $\U5 = U_4 + P_3$				& \bf $C_{12}$ \\  
			10 & $T_4 = T_2 - B_{21}$			& $C_{12}$	& 21 & $P_2 = \ipr(A_{12} B_{21})$		& $B_{12}$ \\
			11 & $P_6 = \ipr(S_2 T_2)$			& $C_{21}$	& 22 & $\U1 = P_1 + P_2$				& \bf $C_{11}$  \\  
			\hline
		\end{tabular}
		\caption{\ipr schedule for operation $C\leftarrow A\times B$ using strictly two
		blocks of $B$ and one temporary}
		\label{tab:AB:ipright}
	\end{center}
\end{table}
We note that we can overwrite only two blocks of $A$ in \ipl when the schedule is modified as follows:
\begin{center}
	\newcolumntype{s}{>{\!\!\scriptsize}l<{\!\!\!\!}}
	\newcolumntype{t}{>{\!\!\!\!\scriptsize}r<{\!\!}}
	\small
	\begin{tabular}{|slt|}
		\hline
		\# & operation & loc. \\
		\hline
		18bis	& $A_{21} = \texttt{Copy}(A_{12})$	& $A_{21}$	\\	
		\hline
		19bis	& $A_{12} = \texttt{Copy}(A_{21})$	& $A_{12}$	\\
		\hline
		21		& $P_4 = \ipr(A_{22} T_4)$			& $A_{21}$	\\ 
		\hline
	\end{tabular}
\end{center}
Similarly, for \ipr, we can overwrite only two blocks of $B$ using copies on lines 20 and 21 and \ipl on line 19.\\
%
%
We now compute the extra memory needed 
for the schedule of table~\ref{tab:AB:ipright}.
The size of the temporary block $X$ is 
$\left(\frac{n}{2}\right)^2$, 
the extra memory required for table~\ref{tab:AB:ipright} hence satisfies:
$E_{\ref{tab:AB:ipright}}(n,n,n) < \frac{1}{3} n^2$.
\subsection{Product with accumulation}
We now consider the operation $C\leftarrow \alpha A\times B + \beta C$,
where the input matrices $A$ and $B$ can be overwritten. We propose in table~\ref{tab:ABC:overwrite} a schedule that only requires $2$ temporary block
matrices, instead of the $3$ in table~\ref{tab:schedule:ABC}. This is
achieved by overwriting the inputs and by using two additional pre-additions ($Z_1$ and $Z_2$) on the matrix $C$.
\arraycolsep 0pt
\begin{table}[htb]
	\scriptsize
	\newcolumntype{s}{>{\!\!\!\!\scriptsize}l<{\!\!\!\!\!\!}}
	\newcolumntype{t}{>{\!\!\!\!\!\!\scriptsize}r<{\!\!\!\!}}
	\begin{center}
		\begin{tabular}{|slt|slt|}
			\hline
			\# & operation & loc. & \# & operation & loc.  \\
			\hline
			1  & $Z_1 = C_{22} - C_{12}$            & $C_{22}$  & 13 & $P_4 = \acco(\alpha A_{22} T_4{-\beta Z_2})$       & $C_{21}$ \\ 
			2  & $S_1 = A_{21} + A_{22}$            & $X$       & 14 & $S_4 = A_{12} - S_2$                         & $A_{22}$ \\  
			3  & $T_1 = B_{12} - B_{11}$            & $Y$       & 15 & $P_6 = \alpha \ip(S_2 T_2)$                  & $X$  \\
			4  & $Z_2 = C_{21} - Z_1$               & $C_{21}$  & 16 & $P_2 = \acco(\alpha A_{12} B_{21}{+\beta C_{11}})$  & $C_{11}$ \\
			5  & $T_3 = B_{22} - B_{12}$            & $B_{12}$  & 17 & $\U1 = P_1 + P_2$                            & $C_{11}$ \\
			6  & $S_3 = A_{11} - A_{21}$            & $A_{21}$  & 18 & $U_2 = P_1 + P_6$                            & $X$  \\
			5  & $P_7 = \acco(\alpha S_3 T_3{+\beta Z_1})$ & $C_{22}$  & 17 & $U_3 = U_2 + P_7$                            & $C_{22}$ \\
			8  & $S_2 = S_1 - A_{11}$               & $A_{21}$  & 20 & $U_4 = U_2 + P_5$                            & $X$  \\
			9  & $T_2 = B_{22} - T_1$               & $B_{12}$  & 21 & $\U6 = U_3 - P_4$                            & $C_{21}$ \\  
			10 & $P_5 = \acco(\alpha S_1 T_1{+\beta C_{12}})$& $C_{12}$  & 22 & $\U7 = U_3 + P_5$                            & $C_{22}$ \\
			11 & $P_1 = \alpha \ip(A_{11} B_{11}) $ & $Y$		& 23 & $P_3 = \alpha \ip(S_4 B_{22})$               & $C_{12}$ \\
			12 & $T_4 = T_2 - B_{21}$               & $X$		& 24 & $\U5 = U_4 + P_3$                            & $C_{12}$ \\
			\hline
		\end{tabular}
		\caption{\acco schedule for $C\leftarrow \alpha A\times B + \beta C$ overwriting~$A$~and~$B$ with 2
		temporaries, 4 recursive calls}
		\label{tab:ABC:overwrite}
	\end{center}
\end{table}
We also propose in table~\ref{tab:ABC:overright} a schedule
similar to table~\ref{tab:ABC:overwrite}
overwriting only for instance the right input matrix. 
It also uses only two temporaries, but has to call the \ipr schedule.
The extra memory required by $X$ and $Y$ in table~\ref{tab:ABC:overwrite} is
$2\left(\frac{n}{2}\right)^2$.
	Hence, using lemma~\ref{lem:sum}:
	\begin{equation}\label{Eq:overwrite}
	E_{\ref{tab:ABC:overwrite}}(n,n,n) < \frac{2}{3}   n^2 .
\end{equation}
%
\begin{table}[htb]
	\scriptsize
	\newcolumntype{s}{>{\!\!\!\!\scriptsize}l<{\!\!\!\!\!\!}}
	\newcolumntype{t}{>{\!\!\!\!\!\!\scriptsize}r<{\!\!\!\!}}
	\begin{center}
		\begin{tabular}{|slt|slt|}
			\hline
			\# & operation & loc. & \# & operation & loc.  \\
			\hline
			1  & $Z_1 = C_{22} - C_{12} $					& $C_{22}$	& 13 & $P_2 = \accr(\alpha A_{12} B_{21}{+\beta C_{11}})$	& $C_{11}$	\\
			2  & $T_1 = B_{12} - B_{11}$					& $X$ 		& 14 & $S_2 = S_1 - A_{11}$									& $Y$		\\
			3  & $Z_2 = C_{21} - Z_1$						& $C_{21}$	& 15 & $P_6 = \alpha \ipr(S_2 T_2)$							& $B_{21}$	\\
			4  & $T_3 = B_{22} - B_{12}$					& $B_{12}$	& 16 & $S_4 = A_{12} - S_2$									& $Y$		\\
			5  & $S_3 =A_{11} - A_{21}$						& $Y$ 		& 17 & $U_2 = P_1 + P_6$									& $B_{21}$  \\
			6  & $P_7 = \accr(\alpha S_3 T_3{+\beta Z_1})$	& $C_{22}$	& 18 & $U_3 = U_2 + P_7$									& $C_{22}$  \\
			7  & $S_1 = A_{21} + A_{22}$					& $Y$		& 19 & $U_4 = U_2 + P_5$									& $B_{21}$	\\
			8  & $T_2 = B_{22} - T_1$						& $B_{12}$	& 20 & $\U6 = U_3 - P_4$									& $C_{21}$  \\
			9  & $P_5 = \accr(\alpha S_1 T_1{+\beta C_{12}})$ & $C_{12}$	& 21 & $\U1 = P_1 + P_2$									& $C_{11}$  \\
			10 & $T_4 = T_2 - B_{21}$						& $X$		& 22 & $\U7 = U_3 + P_5$									& $C_{22}$  \\
			11 & $P_4 = \accr(\alpha A_{22} T_4{-\beta Z_2})$ & $C_{21}$ 	& 23 & $P_3 = \alpha \ip(S_4 B_{22})$					& $C_{12}$  \\
			12 & $P_1 = \alpha \ipr(A_{11} B_{11})$			& $X$ 		& 24 & $\U5 = U_4 + P_3$									& $C_{12}$  \\
			\hline
		\end{tabular}
		\caption{\accr schedule for $C\leftarrow \alpha
                  A\times B + \beta C$ overwriting $B$ with 2
		temporaries, 4
		recursive calls}
		\label{tab:ABC:overright}
	\end{center}
\end{table}
The extra memory $E_{\ref{tab:ABC:overright}}(n,n,n)$ required for table~\ref{tab:ABC:overright} in the top level of recursion is:
$$
\left(\frac{n}{2}\right)^2 + \left(\frac{n}{2}\right)^2 + \max{\left(E_{\ref{tab:ABC:overright}},E_{\ref{tab:AB:ipright}}\right)}\left(\frac{n}{2},\frac{n}{2},\frac{n}{2}\right).
$$
We clearly have $E_{\ref{tab:ABC:overright}}>E_{\ref{tab:AB:ipright}}$ and:  
$$
E_{\ref{tab:ABC:overright}}(n,n,n) < \frac{2}{3}n^2.
$$
 Compared with the schedule of table~\ref{tab:schedule:ABC}, the
 possibility to
 overwrite the input matrices makes it possible to have further in place
 calls and replace recursive calls with accumulation by calls without
 accumulation. We show in theorem~\ref{thm:cost} that this enables us to
 almost compensate for the extra additions performed.

\subsection{The rectangular case}\label{ssec:overwrite:gen}
We now examine the sizes of the temporary locations used, when the
matrices involved do not have identical sizes. We want to make use of
table~\ref{tab:AB:inplace} for the general case.\\
Firstly, the sizes of $A$ and $B$ must not be bigger than that of $C$ (\ie{} we need $k \leq \min{(m,n)}$). Indeed, let's play a pebble game that we start with pebbles on the inputs and  $4$ extra pebbles that are the size of a $C_{ij}$. No initial pebble can be moved since at least two edges initiate from the initial nodes. If the size of $A_{ij}$ is larger that the size of the free pebbles, then we cannot put a free pebble on the $S_i$ nodes (they are too large). We cannot put either a pebble on $P_1$ or $P_2$ since their operands would be overwritten. So the size of $A_{ij}$ is smaller or equal than that of $C_{ij}$. The same reasoning applies for $B_{ij}$.\\ 
Then, if we consider a pebble game that was successful, we can prove in the same fashion that either the size of $A$ or the size of $B$ can not be smaller that of $C$ (so one of them has the same size as $C$).\\
Finally, table~\ref{tab:AB:inplace} shows that this is indeed possible, with
$k=n\leq m$. It is also possible  to switch the roles of $m$ and $n$. \\
Now in tables~4 to~7, we need that $A$, $B$ and $C$ have the same size.
Generalizing table~\ref{tab:AB:inplace} whenever we do not have a dedicated
in-place schedule can then done by cutting the larger matrices in squares of
dimension $\min{(m,k,n)}$ and doing the multiplications / product with accumulations on
these smaller matrices using
algorithm~1 to~7 and free space from $A$, $B$ or $C$.
Since algorithms~1 to~7 require less than $n^2$ extra memory, we can use them as soon as one small matrix is free.\\
We now propose an example in algorithm \ref{alg:ipmm0} for the case $n < \min{(m,k)}$:
\begin{algorithm}[htb] 
	\begin{algorithmic}[1]
		\Require{$A$ and $B$ of resp. sizes $m\times k$ and
		$k\times n$} 
		\Require{$n<\min{(m,k)}$ and $m$, $k$, $n$ powers of $2$.}
		\Ensure{$C = A\times B$}
		\State Let $k_0 = k/n$ and $m_0 = m/n$.
		\State Split $A = \left[ \small
		\begin{array}{c|c|c} 
			A_{1,1} & \dots & A_{1,k_0}\\
			\hline
			\vdots & & \vdots \\
			\hline
			A_{m_0,1} & \dots & A_{m_0,k_0}\\
		\end{array}\right]$, $B  = \left[ \small
		\begin{array}{c} 
			B_{1} \\ 
			\hline  \vdots  \\ 
			\hline B_{k_0} 
		\end{array}\right]$ and $C  = \left[ \small
		\begin{array}{c} 
			C_{1} \\ 
			\hline  \vdots  \\ 
			\hline C_{k_0} 
		\end{array}\right]$
		\Comment{
		\begin{minipage}{32mm}
			where $A_{i,j}$ and $B_{j}$ have dimension $n \times n$
		\end{minipage}
		}
		\State $C_{1} \gets A_{1,1}B_{1}$ \Comment{with alg. of table~\ref{tab:schedule:AB} and memory $C_{2}$.}
		\State Now we use  $A_{1,1}$ as temporary space.
		\For {$i=2\dots k_0$}
		\State $C_{i} \gets A_{i,1}B_{1}$ \Comment{with alg. of table~\ref{tab:AB:ipleft}.} 
		\EndFor
		\For {$j=2\dots k_0$}
		\For {$i=1\dots m_0$} 
		\State $C_j \gets  A_{i,j}B_{j} + C_j$ \Comment{
			with alg. of table~\ref{tab:schedule:ABC}.
		}
		\EndFor
		\EndFor
	\end{algorithmic}
	\caption{\texttt{IP0vMM}: In-Place Overwrite Matrix Multiply}\label{alg:ipmm0}
\end{algorithm}
\begin{prop}
	Algorithm~\ref{alg:ipmm0} computes the product $C = A B$ in place, overwriting $A$ and $B$.
\end{prop}%
%
Finally, we generalize the accumulation operation from
table~\ref{tab:ABC:overright} to the rectangular case. We can no
longer use dedicated square algorithms. This is done in
table~\ref{tab:ABC:overright:gen}, overwriting only one of the inputs
and using only two temporaries, but with 5 recursive accumulation calls: 
\begin{table}[htb]
	\scriptsize
	\newcolumntype{s}{>{\!\!\!\!\scriptsize}l<{\!\!\!\!\!\!}}
	\newcolumntype{t}{>{\!\!\!\!\!\!\scriptsize}r<{\!\!\!\!}}
	\begin{center}
		\begin{tabular}{|slt|slt|}
			\hline
			\# & operation & loc. & \# & operation & loc.  \\
			\hline
			1  & $Z_1 = C_{22} - C_{12} $			& $C_{22}$	& 13 & $P_2 = \accrb(\alpha A_{12} B_{21}{+\beta C_{11}})$	& $C_{11}$	\\
			2  & $T_1 = B_{12} - B_{11}$			& $X$ 		& 14 & $\U1 = P_1 + P_2$							& $C_{11}$  \\
			3  & $Z_2 = C_{21} - Z_1$				& $C_{21}$	& 15 & $S_2 = S_1 - A_{11}$							& $Y$		\\
			4  & $T_3 = B_{22} - B_{12}$			& $B_{12}$	& 16 & $U_2 = \accrb(\alpha S_2 T_2{+P_1})$				    & $X$		\\
			5  & $S_3 = A_{11} - A_{21}$			& $Y$ 		& 17 & $U_3 = U_2 + P_7$							& $C_{22}$		\\
			6  & $P_7 = \accrb(\alpha S_3 T_3{+\beta Z_1})$ & $C_{22}$	& 18 & $\U6 = U_3 - P_4$							& $C_{21}$  \\
			7  & $S_1 = A_{21} + A_{22}$			& $Y$		& 19 & $\U7 = U_3 + P_5$							& $C_{22}$  \\
			8  & $T_2 = B_{22} - T_1$				& $B_{12}$	& 20 & $U_4 = U_2 + P_5$							& $X$		\\
			9  & $P_5 = \accrb(\alpha S_1 T_1{+\beta C_{12}})$& $C_{12}$	& 21 & $S_4 = A_{12} - S_2$							& $Y$		\\ 
			10 & $T_4 = T_2 - B_{21}$				& $X$		& 22 & $P_3 = \alpha S_4 B_{22}$					& $C_{12}$  \\
			11 & $P_4 = \accrb(\alpha A_{22} T_4{-\beta Z_2})$& $C_{21}$ 	& 23 & $\U5 = U_4 + P_3$							& $C_{12}$  \\ 
			12 & $P_1 = \alpha A_{11} B_{11}$		& $X$ 		& 24 & & \\
			\hline
		\end{tabular}
		\caption{\accrb schedule for $C\leftarrow \alpha A\times B + \beta C$ with 5
		recursive calls, 2 temporaries and overwriting $B$}
		\label{tab:ABC:overright:gen}
	\end{center}
\end{table}

For instance, in table \ref{tab:ABC:overright:gen}, the last
multiplication (line 22,  $P_3 = \alpha S_4 B_{22}$) could have been
made by a call to the in place algorithm, would $C_{12}$ be large
enough. This is not always the case in a rectangular setting. 

Now, the size of the extra temporaries required in table
\ref{tab:ABC:overright:gen} is
$\max{\left(\frac{m}{2},\frac{k}{2}\right)}\frac{n}{2}+
\frac{m}{2}\frac{k}{2}$ and  $E_{\ref{tab:ABC:overright:gen}}(m,k,n)$
is equal to: 
$$
\max{\left(\frac{m}{2},\frac{k}{2}\right)}\frac{n}{2}  + \frac{m}{2}\frac{k}{2} + \max{\left(E_{\ref{tab:ABC:overright:gen}},E_{\ref{tab:schedule:AB}}\right)}\left(\frac{m}{2},\frac{k}{2},\frac{n}{2}\right).
$$
If $m<k<n$ or $k < m < n$, then $E_{\ref{tab:ABC:overright:gen}}(m,k,n) < E_{\ref{tab:schedule:AB}}(m,k,n)$: 
\begin{align*}
	E_{\ref{tab:ABC:overright:gen}}(m,k,n) &= \max{\left(\frac{m}{2},\frac{k}{2}\right)}\frac{n}{2}  + \frac{m}{2}\frac{k}{2} + E_{\ref{tab:schedule:AB}}\left(\frac{m}{2},\frac{k}{2},\frac{n}{2}\right)\\
	& <  \max{\left(\frac{m}{2},\frac{k}{2}\right)}\frac{n}{2} + \frac{m}{2}\frac{k}{2} + \frac{1}{3}\left(\frac{m}{2}\frac{n}{2}+ \frac{k}{2}\frac{n}{2}\right). 
\end{align*}
Otherwise $E_{\ref{tab:ABC:overright:gen}}(m,k,n) \geq E_{\ref{tab:schedule:AB}}(m,k,n)$ and:
$$
E_{\ref{tab:ABC:overright:gen}}(m,k,n) < \frac{1}{3}\left(\max{\left(m,k\right)}n  + mk \right).
$$
In the square case, this simplifies into
$E_{\ref{tab:ABC:overright:gen}}(n,n,n) \leq \frac{2}{3}n^2.$\\

In addition, if the size of $B$ is bigger than that of $A$, then  one
can store $S_2$, for instance within $B_{12}$, and separate the
recursive call $16$ into a multiplication and an addition, which
reduces the arithmetic complexity. Otherwise, a scheduling with only 4
recursive calls exists too, but we need for instance to recompute
$S_4$ at step $21$.

\section{Hybrid scheduling}\label{sec:hyb}
By combining techniques from sections~\ref{ssec:constinput} and
\ref{sec:overwrite}, we now propose in table~\ref{tab:ABC:2tmp} a hybrid algorithm that performs the
computation
$C\leftarrow \alpha A\times B + \beta C$ with constant input matrices $A$ and
$B$, with a lower extra memory requirement than the scheduling of
~\cite{Huss-Lederman:1996:mai} (table~\ref{tab:schedule:ABC}).
We have to pay a price of order $n^2\log(n)$ extra
operations, as we need to compute the temporary variable $T_2$ twice.
%
\begin{table}[htb]
	\newcolumntype{s}{>{\!\!\!\!\scriptsize}l<{\!\!\!\!\!\!}}
	\newcolumntype{t}{>{\!\!\!\!\!\!\scriptsize}r<{\!\!\!\!}}
\scriptsize
\begin{center}
\begin{tabular}{|slt|slt|}
\hline
\# & operation & loc. & \# & operation & loc.  \\
\hline
1  & $Z_1 = C_{22} - C_{12}$			& $C_{22}$	& 14 & $P_2 =\acc(\alpha A_{12}B_{21}{+\beta C_{11}})$	& $C_{11}$\\ 
2  & $Z_3 = C_{12} - C_{21}$			& $C_{12}$	& 15 & $\U1 = P_1 + P_2$							& $C_{11}$\\
3  & $S_1 = A_{21} + A_{22}$			& $X$		& 16 & $\U5 = U_2 + P_3$						& $C_{12}$ \\ 
4  & $T_1 = B_{12} - B_{11}$			& $Y$		& 17 & $S_3 = A_{11} - A_{21}$					& $X$\\
5  & $P_5 = \acc(\alpha S_1 T_1{+\beta Z_3})$	& $C_{12}$	& 18 & $T_3 = B_{22} - B_{12}$					& $Y$ \\
6  & $S_2 = S_1 - A_{11}$				& $X$ 		& 19 & $U_3 = P_7 + U_2$						& $C_{21}$ \\
7  & $T_2= B_{22} - T_1$				& $Y$		&    & $\phantom{U_3} = \alpha \acco(S_3 T_3{+U_2})\hspace{-5pt}$&   \\
8  & $P_6 = \acc(\alpha S_2 T_2{+\beta C_{21}})$& $C_{21}$	& 20 & $\U7 = U_3 + W_1$						& $C_{22}$    \\
9  & $S_4 = A_{12} - S_2$				& $X$ 		& 21 & $T_1' = B_{12} - B_{11}$					& $Y$ \\
10 & $W_1= P_5+ \beta Z_1$				& $C_{22}$	& 22 & $T_2' = B_{22} - T_1'$						& $Y$ \\
11 & $P_3 = \acc(\alpha S_4 B_{22}{+P_5})$	& $C_{12}$	& 23 & $T_4 = T_2' - B_{21}$						& $Y$ \\
12 & $P_1 = \alpha A_{11} B_{11}$		& $X$   	& 24 & $\U6 = U_3 - P_4 $						&  $C_{21}$    \\
13 & $U_2 = P_6 + P_1 $					& $C_{21}$  &	 & $\phantom{\U6} = -\alpha \accr(A_{22}T_4{-U_3})\hspace{-15pt}$ & \\
\hline
\end{tabular}
\caption{\acc schedule for operation $C\leftarrow \alpha A\times B+\beta C$ with 2 temporaries}
\label{tab:ABC:2tmp}
\end{center}
\end{table}

Again, the two temporary blocks $X$ and $Y$ 
have dimensions
$X_s = Y_s = (n/2)^2$ so that:
$$E_{\ref{tab:ABC:2tmp}} =Y_s + \max{\left\{X_s + E_{\ref{tab:ABC:2tmp}} ,X_s + E_{\ref{tab:ABC:overwrite}},E_{\ref{tab:ABC:overright:gen}}\right\}} \left(\frac{m}{2}, \frac{k}{2}, \frac{n}{2}\right).$$
In all cases, $E_{\ref{tab:ABC:overwrite}}+X_s \geq
E_{\ref{tab:ABC:overright:gen}}.$ But $X_s + Y_s$ is not as large as
the size of the two temporaries in table~\ref{tab:ABC:overwrite}. We
therefore get:
\begin{align*}
	E_{\ref{tab:ABC:2tmp}}(m,k,n) = {} & Y_s + X_s + E_{\ref{tab:ABC:overwrite}}\left(\frac{m}{2}, \frac{k}{2}, \frac{n}{2}\right)\\
						   < {} & 2 \left(\frac{n}{2}\right)^2   +  \frac{1}{3} \left( \left(\frac{n}{2}\right)^2 + \left(\frac{n}{2} \right)^2\right).
\end{align*}

Assuming $m=n=k$, one gets  $E_{\ref{tab:ABC:2tmp}}(n,n,n) <\frac{2}{3}n^2,$ which is smaller than the extra memory requirement of table \ref{tab:schedule:ABC}.\\

%
%
\section{A sub-cubic in-place algorithm}\label{sec:mix}
Following the improvements of the previous section, the question was raised
whether extra memory allocation was intrinsic to sub-cubic matrix multiplication
algorithms. More precisely, is there a matrix multiplication algorithm computing
$C\leftarrow A\times B$ in
$\GO{n^{\log_2 7}}$ arithmetic operations without extra memory allocation and without
overwriting its input arguments? We show in this section that a combination of Winograd's algorithm and a classic block algorithm provides a positive answer.
Furthermore this algorithm also improves the extra memory requirement for the
product with accumulation $C\leftarrow \alpha A\times B + \beta C$.
%
%
\subsection{The algorithm}\label{sec:fully}
The key idea is to split the result matrix $C$ into four quadrants of dimension
$n/2 \times n/2$. The first
three quadrants $C_{11}, C_{12}$ and $C_{21}$ are computed using fast
rectangular matrix multiplication, which accounts for $2k/n$ standard
Winograd multiplications on blocks of dimension $n/2 \times n/2$. The temporary
memory for these computations is stored in $C_{22}$. Lastly, the block $C_{22}$
is  computed recursively up to a base case,
as shown on algorithm~\ref{alg:ipmm}. This base case, when the matrix
is too small to benefit from the fast routine, is then computed
with the classical matrix multiplication.
\begin{algorithm}[htb]
	\algblock{Do}{EndDo}
	\algrenewtext{Do}{\textbf{do}}
	\algrenewtext{EndDo}{\textbf{end do}}
	\begin{algorithmic}[1]
		\Require{$A$ and $B$, of dimensions resp. $n\times k$ and $k\times n$ with
		$k$,  $n$ powers of 2 and $k\geq n$.}
		\Ensure{$C = A\times B$}
		\State Split $C=\begin{bmatrix}
			C_{11}&C_{12}\\
			C_{21}&C_{22}
		\end{bmatrix}$, $A = \left[\small
		\begin{array}{c|c|c} 
			A_{1,1} & \dots & A_{1,2k/n}\\
			\hline A_{2,1}&\dots& A_{2,2k/n}
		\end{array}\right]$ and
		$B  = \left[\small
		\begin{array}{c|c}
			B_{1,1}&B_{1,2} \\ 
			\hline  \vdots & \vdots\\ 
			\hline B_{2k/n,1} &  B_{2k/n,2} 
		\end{array}
		\right]$	\Comment \begin{minipage}{3cm}
			where each $A_{i,j}, B_{i,j}$ and $C_{i,j}$ have dimension $n/2 \times n/2$.
		\end{minipage}
		\Do \Comment{with alg. of table~\ref{tab:schedule:AB} using $C_{22}$ as temp. space}
		\State $C_{11}=A_{1,1}B_{1,1}$
		\State $C_{12}=A_{1,1}B_{1,2}$
		\State $C_{21}=A_{2,1}B_{1,1}$
		\EndDo
		\For {$i=2\dots \frac{2k}{n}$} \Comment{with alg. of table~\ref{tab:schedule:ABC} using $C_{22}$ as temporary space:}
		\State $C_{11} =  A_{1,i}B_{i,1} + C_{11}$
		\State $C_{12} =  A_{1,i}B_{i,2} + C_{12}$
		\State $C_{21} =  A_{2,i}B_{i,1} + C_{21}$ 
		\EndFor
		\State $C_{22} = A_{2,*}\times B_{*,2}$ \Comment{recursively using \texttt{IPMM}.}
	\end{algorithmic}
	\caption{\texttt{IPMM}: In-Place Matrix Multiply}\label{alg:ipmm}
\end{algorithm}
%
%
\begin{thm}
The complexity of algorithm~\ref{alg:ipmm} is:
$$G(n,n) = 7.2 n^{\log_2(7)}-13n^2+6.8n$$ when $k=n$.
\end{thm}
\begin{proof}
Recall that  the cost of Winograd's algorithm for square matrices is $W(n) =
6n^{\log_2 7} - 5n^2$  for the operation $C\leftarrow A\times B$ and $W_\text{acc}(n) =
6n^{\log_2 7} - 4n^2$ for the operation $C\leftarrow A\times B + C$.
The cost $G(n,k)$ of algorithm~\ref{alg:ipmm} is given by the relation
$$G(n,k) = 3 W(n/2) +3(2k/n-1)W_\text{acc}(n/2)+G(n/2,k),$$ the base case
being a classical dot product: $G(1,k)=2k-1$.
Thus, $G(n,k) = 7.2 kn^{\log_2(7)-1}-12 kn-n^2+34 k / 5$.
\end{proof}
\begin{thm}
For any $m$, $n$ and $k$, algorithm~\ref{alg:ipmm} is in place.
\end{thm}
\begin{proof}
W.l.o.g, we assume that $m \geq n > 1$ (otherwise we could use the transpose).
	The exact amount of extra memory from algorithms in table~\ref{tab:schedule:AB} and~\ref{tab:schedule:ABC} is
	respectively given by eq.~(\ref{Eq:schedule:AB}) and~(\ref{Eq:schedule:ABC}).\\ 
	If we cut $B$ into $p_i$ stripes at recursion level $i$, then the sizes for the involved submatrices of $A$ (resp. $B$) are $m/2^i \times k/p_i$ (reps. $k/p_i \times n/2^i$). 
	The lower right corner submatrix of $C$ that we would like to use as temporary space has a size $m/2^i \times n/2^i$.
	Thus we need to ensure that the following inequality holds:
	\begin{equation} \label{Eq:aim}
		\max{(E_{\ref{tab:schedule:AB}},E_{\ref{tab:schedule:ABC}})} \left( \frac{m}{2^i}, \frac{k}{p_i}, \frac{n}{2^i} \right) \leq \frac{m}{2^i}  \frac{n}{2^i}.
	\end{equation}
	It is clear that $E_{\ref{tab:schedule:AB}}<E_{\ref{tab:schedule:ABC}},$ which simplifies the previous inequality.
	Let us now write $K = k/p_i$, $M = m/2^i$ and $N = n/2^i$. We need to find, for every $i$ an integer $p_i>1$ so that eq.~(\ref{Eq:aim}) holds. In other words, let us show that there exists some $K<k$ such that, for any $(M,N)$, the inequality $E_{\ref{tab:schedule:ABC}}(M,K,N) \leq M N$ holds.
	Then the fact that $E(M,2,N) < \frac{1}{3} (2M+2N+MN)
        \leq \frac{1}{3}(4M+MN) \leq MN$ provides at least one such $K$.\\
	As the requirements in algorithm~\ref{alg:ipmm} ensure that $k>N$ and $M=N$, there just remains to prove that $E(M,N,N) \leq MN$. Since $E(M,N,N)<\frac{1}{3}(2MN+N^2)$ and  again $M \geq N$, algorithm~\ref{alg:ipmm} is indeed in place.
\end{proof}
Hence a fully in-place $\GO{n^{\log_2 7}}$ algorithm is obtained for matrix
multiplication.
The overhead of this approach appears in the multiplicative constant of
the leading term of the complexity, growing from $6$ to $7.2$.\\
This approach extends to the case of matrices with general
dimensions, using for instance peeling or padding techniques.\\
%
It is also useful if any sub-cubic algorithm is used
instead of Winograd's. For instance, in the square case, one can use the product with accumulation in table~\ref{tab:ABC:2tmp} instead of table~\ref{tab:schedule:ABC}.
\subsection{Reduced memory usage for the product with accumulation}\label{sec:red}
In the case of computing the product with accumulation, the matrix $C$ can no longer be used as temporary storage, and
extra memory allocation cannot be avoided.
Again we can use the idea of the classical block matrix multiplication at the
higher level and call Winograd algorithm for the block multiplications.
As in the previous subsection, $C$ can be divided into
four blocks and then the product can be made with 8 calls to Winograd algorithm for
the smaller blocks, with only one extra temporary block of  dimension $n/2
\times n/2$.\\
More generally, for square $n \times n$ matrices,
$C$ can be divided in $t^2$ blocks of dimension $ \frac{n}{t} \times
\frac{n}{t}$.
Then one can compute each block with Winograd algorithm using only
one extra memory chunk of size $(n/t)^2$. The complexity is changed to
$R_t(n) = t^2 t W_\text{acc}(n/t),$ which is
$R_t(n) = 6t^{3-\log_2(7)}n^{\log_2(7)}-4tn^2$
for an accumulation product with
Winograd's algorithm.
Using the parameter $t$, one can then balance the memory usage
and the extra arithmetic operations. 
For example, with $t=2$,
$$R_2 = 6.857n^{\log_2 7}-8n^2~~\text{and}~~\text{ExtraMem}=\frac{n^2}{4}$$ and
with $t=3$,
$$R_3 = 7.414n^{\log_2 7}-12n^2~~\text{and}~~\text{ExtraMem}=\frac{n^2}{9}.$$
Note that one can use the algorithm of table~\ref{tab:ABC:2tmp} instead of
the classical Winograd accumulation as the base case algorithm.
Then the memory overhead drops down to $\frac{2n^2}{3t^2}$ and the
arithmetic complexity increases to
$R_t(n)+t^{2-\log_2(3)}n^{\log_2(6)}-tn^2$.
%
\section{Conclusion}
With constant input matrices, we reduced the number of extra memory
allocations for the operation $C\leftarrow \alpha A\times B +\beta C$
from $n^2$ to $\frac{2}{3} n^2$,
by introducing two extra pre-additions. As shown below, the
overhead induced by these supplementary additions is amortized by
the gains in number of memory allocations.

If the input matrices can be overwritten, we proposed a fully \textit{in-place}
schedule for the operation $C\leftarrow A\times B$ without any extra operations.
We also proposed variants for the operation $C\leftarrow
A\times B$, where only
one of the input matrices is being overwritten and one temporary is
required.
These subroutines allow us to reduce the extra memory allocations
required for the $C\leftarrow \alpha
A\times B +\beta C$ operation without overwrite: the extra required
temporary space drops from $n^2$ to only $\frac{2}{3} n^2$, at a negligible cost.

Some algorithms with an even more reduced memory usage, but with some increase in arithmetic
complexity, are also shown.
Table~\ref{tab:resume} gives a summary of the features of each schedule that has been
presented.
The complexities are given only for $m=k=n$ being a power of $2$.
%

%
\begin{sidewaystable}[!htbp]
\small
\begin{center}
	\newcolumntype{r}{>{\centering}m{1.7cm}}
	\newcolumntype{s}{>{\centering}m{30pt}}
	\newcolumntype{t}{>{\centering}m{3.3cm}}
	\begin{tabular}{|c|c|c|r|s|t|c|}
\hline
& Algorithm & Input matrices    &\hspace{-10pt} \# of extra temporaries \hspace{-10pt}&total extra memory& total \# of extra allocations & arithmetic complexity\\
\hline
& & & & & &\\[-5pt]
\multirow{4}{*}{\begin{sideways}\footnotesize$ A \times B $\end{sideways}}& Table~\ref{tab:schedule:AB} \dhss & Constant & $2$  & $\frac{2}{3}n^2$   & $\frac{2}{3}(n^{2.807}-n^2)$ & $6n^{2.807}-5n^2$\\
& Table~\ref{tab:AB:inplace}							& Both Overwritten		   & $0$  &    $0$				& $0$ &$6n^{2.807}-5n^2$\\
& Table~\ref{tab:AB:ipleft} or~\ref{tab:AB:ipright}     & $A$ or $B$ Overwritten   &  $1$ & $\frac{1}{3}n^2$    & $\frac{1}{4}n^2\log_2(n)$& $6n^{2.807}-5n^2$\\
& \ref{sec:fully}                                       & Constant                 &  $0$ &     $0$				& $0$ & $7.2n^{2.807}-13n^2$\\[1 ex]
\hline 
& & & & & &\\[-5pt]
\multirow{6}{*}{\begin{sideways}\footnotesize$\alpha A\times B+\beta
    C$\end{sideways}} & Table~\ref{tab:schedule:ABC} \hljjtt &
Constant & $3$  &     $n^2$ &
$\frac{2}{3}n^{\log_2(7)}+n^{\log_2(5)}-\frac{5}{3}n^2$ & $6n^{2.807}-4n^2$ \\
& Table~\ref{tab:ABC:overwrite}       & Both Overwritten    & $2$ &$\frac{2}{3}n^2$ & $\frac{1}{2}n^{2}\log_2(n)$ & $6n^{2.807}-4n^2+\frac{1}{2}n^{2}\log_2(n)$\\
& Table~\ref{tab:ABC:overright}        & $B$    Overwritten    & $2$ & $\frac{2}{3}n^2$  & $2n^{2.322}-2n^2$  & $6n^{2.807}-4n^2+\frac{1}{2}n^{2}\log_2(n)$\\
& Table~\ref{tab:ABC:2tmp}        & Constant       & $2$ &$\frac{2}{3}n^2$  & $\frac{2}{9}n^{2.807}+2n^{2.322}-\frac{22}{9}n^2$ & $6n^{2.807}-4n^2+\frac{4}{3}n^2\log_2(n)$\\
&~\ref{sec:red} & Constant & N/A & $\frac{1}{4} n^2$ & $\frac{1}{4} n^2$ & $6.857n^{2.807}-8n^2$\\
&~\ref{sec:red} & Constant & N/A & $\frac{1}{9} n^2$ & $\frac{1}{9} n^2$ & $7.414n^{2.807}-12n^2$\\[1 ex]
\hline
\end{tabular}
\caption{Complexities of the schedules presented for square matrix multiplication}
\label{tab:resume}
\end{center}
\end{sidewaystable}
%
\begin{thm}\label{thm:cost}
The arithmetic and memory complexities of table~\ref{tab:resume} are
correct.
\end{thm}
\begin{proof}
For the operation $A\times B$, the arithmetic complexity of the schedule of
table~\ref{tab:schedule:AB} classically satisfies
$$\left\{\begin{array}{rcl}
W_{\ref{tab:schedule:AB}}(n) &=& 7W_{\ref{tab:schedule:AB}}(\frac{n}{2})+15\left(\frac{n}{2}\right)^2\\
W_{\ref{tab:schedule:AB}}(1)&=&1
\end{array}\right.
,$$
so that $W_{\ref{tab:schedule:AB}}(n) = 6n^{\log_2(7)}-5n^2$.

The schedule of table~\ref{tab:schedule:AB} requires
$$
\left\{\begin{array}{rcl}
M_{\ref{tab:schedule:AB}}(n)&=&2\left(\frac{n}{2}\right)^2 +
       M_{\ref{tab:schedule:AB}}\left(\frac{n}{2}\right)\\
M_{\ref{tab:schedule:AB}}(1)&=&0
\end{array}\right.$$
extra memory space, which is $M_{\ref{tab:schedule:AB}}(n) = \frac{2}{3}n^2$. Its total
number of allocations satisfies
$
A_{\ref{tab:schedule:AB}}(n) = 2\left(\frac{n}{2}\right)^2 +
7A_{\ref{tab:schedule:AB}}\left(\frac{n}{2}\right)
$
which is $A_{\ref{tab:schedule:AB}}(n)=\frac{2}{3}(n^{\log_2(7)}-n^2)$.

The schedule of table~\ref{tab:AB:ipleft} requires
$M_{\ref{tab:AB:ipleft}}(n) =
\left(\frac{n}{2}\right)^2+M_{\ref{tab:AB:ipleft}}\left(\frac{n}{2}\right)$
extra memory space, which is $M_{\ref{tab:AB:ipleft}}(n) = \frac{1}{3}n^2$. Its
total number of allocations satisfies
$
A_{\ref{tab:AB:ipleft}}(n) = \left(\frac{n}{2}\right)^2 +
4A_{\ref{tab:AB:ipleft}}\left(\frac{n}{2}\right)
$
which is $A_{\ref{tab:AB:ipleft}}(n) = \frac{1}{4}n^2\log_2(n)$.

The schedule of table~\ref{tab:AB:ipright} requires the same amount of
arithmetic operations or memory.

For $A\times B + \beta C$, the arithmetic complexity of \hljjtt satisfies
$$W_{\ref{tab:schedule:ABC}}(n)=5W_{\ref{tab:schedule:ABC}}\left(\frac{n}{2}\right)+2W_{\ref{tab:schedule:AB}}\left(\frac{n}{2}\right)+14\left(\frac{n}{2}\right)^2,$$
hence $W_{\ref{tab:schedule:ABC}}(n)=6n^{\log_2(7)}-4n^2$; its memory overhead satisfies
$
M_{\ref{tab:schedule:ABC}}(n)=3\left(\frac{n}{2}\right)^2 +
  M_{\ref{tab:schedule:ABC}}\left(\frac{n}{2}\right),
$
which is $M_{\ref{tab:schedule:ABC}}(n)=n^2$; its total number of allocations satisfies
$
A_{\ref{tab:schedule:ABC}}(n)=3\left(\frac{n}{2}\right)^2 +
5A_{\ref{tab:schedule:ABC}}\left(\frac{n}{2}\right)+2A_{\ref{tab:schedule:AB}}\left(\frac{n}{2}\right),
$
which is
$$
A_{\ref{tab:schedule:ABC}}(n)=\frac{2}{3}n^{\log_2(7)}+n^{\log_2(5)}-\frac{5}{3}n^2.
$$

The arithmetic complexity of the schedule of table~\ref{tab:ABC:overwrite} satisfies
$$W_{\ref{tab:ABC:overwrite}}(n) = 4
W_{\ref{tab:ABC:overwrite}}\left(\frac{n}{2}\right) +
3W_{\ref{tab:schedule:AB}}\left(\frac{n}{2}\right)+17\left(\frac{n}{2}\right)^2,
$$ so that
$W_{\ref{tab:ABC:overwrite}}(n) = 6n^{\log_2(7)}-4n^2+\frac{1}{2}n^{2}\log_2(n)$;
its number of extra memory satisfies
$
M_{\ref{tab:ABC:overwrite}}(n) = 2\left(\frac{n}{2}\right)^2 +
M_{\ref{tab:ABC:overwrite}}\left(\frac{n}{2}\right),
$ which is $M_{\ref{tab:ABC:overwrite}}(n)=\frac{2}{3}n^2$;
its total number of allocations satisfies
$A_{\ref{tab:ABC:overwrite}}(n) = 2\left(\frac{n}{2}\right)^2 +
4A_{\ref{tab:ABC:overwrite}}\left(\frac{n}{2}\right),
$ which is
$A_{\ref{tab:ABC:overwrite}}(n)=n^2+\frac{1}{2}n^{2}\log_2(n)$.

The arithmetic complexity of table~\ref{tab:ABC:overright} schedule satisfies
$$W_{\ref{tab:ABC:overright}}(n) = 4
W_{\ref{tab:ABC:overright}}\left(\frac{n}{2}\right) +
W_{\ref{tab:schedule:AB}}\left(\frac{n}{2}\right) +
2W_{\ref{tab:AB:ipright}}\left(\frac{n}{2}\right) +
16\left(\frac{n}{2}\right)^2,
$$ so that
$W_{\ref{tab:ABC:overright}}(n) = 6n^{\log_2(7)}-4n^2+\frac{1}{2}n^{2}\log_2(n)$;
its number of extra memory satisfies
$
M_{\ref{tab:ABC:overright}}(n) = 2\left(\frac{n}{2}\right)^2 +
M_{\ref{tab:ABC:overright}}\left(\frac{n}{2}\right),
$ which is $M_{\ref{tab:ABC:overright}}(n)=\frac{2}{3}n^2$;
its total number of allocations satisfies
$A_{\ref{tab:ABC:overright}}(n) = 2\left(\frac{n}{2}\right)^2 +
4A_{\ref{tab:ABC:overright}}\left(\frac{n}{2}\right) +
2A_{\ref{tab:AB:ipright}}\left(\frac{n}{2}\right),
$ which is
$A_{\ref{tab:ABC:overright}}(n)=2n^{\log_2(5)}-2n^2$.

The arithmetic complexity of the schedule of table ~\ref{tab:ABC:2tmp} satisfies
$$
W_{\ref{tab:ABC:2tmp}}(n)=4W_{\ref{tab:ABC:2tmp}}\left(\frac{n}{2}\right) +
W_{\ref{tab:schedule:AB}}\left(\frac{n}{2}\right) +
2W_{\ref{tab:ABC:overwrite}}\left(\frac{n}{2}\right) +
17\left(\frac{n}{2}\right)^2,
$$ so that
$W_{\ref{tab:ABC:2tmp}}(n)=6n^{\log_2(7)}-4n^2+\frac{4}{3}n^2\left(\log_2(n)-\frac{10}{3}\right)+\frac{4}{9}$;
its number of extra memory satisfies
$
M_{\ref{tab:ABC:2tmp}}(n) = 2\left(\frac{n}{2}\right)^2 +
M_{\ref{tab:ABC:2tmp}}\left(\frac{n}{2}\right),
$
which is $M_{\ref{tab:ABC:2tmp}}(n)=\frac{2}{3}n^2$;
its total number of allocations satisfies
$
A_{\ref{tab:ABC:2tmp}}(n) = 2\left(\frac{n}{2}\right)^2 + 4A_{\ref{tab:ABC:2tmp}}\left(\frac{n}{2}\right)+A_{\ref{tab:schedule:AB}}\left(\frac{n}{2}\right)+2A_{\ref{tab:ABC:overwrite}}\left(\frac{n}{2}\right),
$
which is $A_{\ref{tab:ABC:2tmp}}(n)=\frac{2}{9}n^{\log_2(7)}+2n^{\log_2(5)}-\frac{22}{9}n^2+\frac{2}{9}$.
\end{proof}
For instance, by adding up allocations and arithmetic operations in table~\ref{tab:resume},
one sees that the overhead in arithmetic operations of the schedule of table~\ref{tab:ABC:2tmp}
is somehow amortized by the decrease of memory allocations. Thus it makes it
theoretically competitive with the algorithm
of~\cite{Huss-Lederman:1996:mai} as soon as $n > 44$.

Also, problems with dimensions that are not powers of two can be handled 
by combining the cuttings of algorithms \ref{alg:ipmm0} and
\ref{alg:ipmm} with peeling or padding techniques. Moreover,
some cut-off can be set in order to stop the recursion and switch
to the classical algorithm. The use of these cut-offs
will in general decrease both the extra memory requirements and the arithmetic
complexity overhead.

For instance we show on table \ref{tab:ipmmperf} the relative speed of
different multiplication procedures for some double floating point 
rectangular matrices. We use atlas-3.9.4 for the BLAS and a cut-off
of 1024.
We see that pour new schedules perform quite competitively with the
previous ones and that the savings in memory enable larger
computations (MT for memory thrashing).
\begin{table}[htbp]\center
\begin{tabular}{|c||r|r|r|r|}
\hline
Dims. $(m,k,n)$ & Classic & \dhss{} & \texttt{IPMM} & \texttt{IP0vMM} \\
\hline
(4096,4096,4096)  & 14.03  & 11.93   & 13.59  & 11.98\\
(4096,8192,4096)  & 28.29  & 23.39   & 27.16  & 23.88  \\
(8192,8192,8192)  & 113.07 & 85.97   & 98.75  & 85.02\\
(8192,16384,8192) & 231.86 & MT      & 197.24 & 170.72\\
\hline
\end{tabular}
\caption{Rectangular matrix multiplication: computation time in seconds  on a core2 duo,
  3.00GHz, $2\times$2Gb RAM}\label{tab:ipmmperf}
\end{table}


\begin{thebibliography}{10}

\bibitem{Bader:2006:comm}
M.~Bader and C.~Zenger.
\newblock Cache oblivious matrix multiplication using an element ordering based
  on a {Peano} curve.
\newblock {\em Linear Algebra and its Applications}, 417(2--3):301--313, Sept.
  2006.

\bibitem{bailey:603}
D.~H. Bailey.
\newblock Extra high speed matrix multiplication on the {C}ray-$2$.
\newblock {\em SIAM Journal on Scientific and Statistical Computing},
  9(3):603--607, 1988.

\bibitem{Bini:1994:PMC-FA}
D.~Bini and V.~Pan.
\newblock {\em Polynomial and Matrix Computations, Volume 1: Fundamental
  Algorithms.}
\newblock Birkhauser, Boston, 1994.

\bibitem{burgisser:1997}
M.~Clausen, P.~{B\"urgisser}, and M.~A. Shokrollahi.
\newblock {\em Algebraic Complexity Theory}.
\newblock Springer, 1997.

\bibitem{Coppersmith:1990:MMAP}
D.~Coppersmith and S.~Winograd.
\newblock Matrix multiplication via arithmetic progressions.
\newblock {\em Journal of Symbolic Computation}, 9(3):251--280, 1990.

\bibitem{Douglas:1994:gemmw}
C.~C. Douglas, M.~Heroux, G.~Slishman, and R.~M. Smith.
\newblock {GEMMW}: A portable level~3 {BLAS} {Winograd} variant of {Strassen}'s
  matrix-matrix multiply algorithm.
\newblock {\em Journal of Computational Physics}, 110:1--10, 1994.

\bibitem{Dumas:2002:FFLAS}
J.-G. Dumas, T.~Gautier, and C.~Pernet.
\newblock Finite field linear algebra subroutines.
\newblock In T.~Mora, editor, {\em {ISSAC}'2002}, pages 63--74. ACM Press, New
  York, July 2002.

\bibitem{Dumas:2004:FFPACK}
J.-G. Dumas, P.~Giorgi, and C.~Pernet.
\newblock {FFPACK}: Finite field linear algebra package.
\newblock In J.~Gutierrez, editor, {\em {ISSAC}'2004}, pages 119--126. ACM
  Press, New York, July 2004.

\bibitem{Huss-Lederman:1996:ISA}
S.~Huss-Lederman, E.~M. Jacobson, J.~R. Johnson, A.~Tsao, and T.~Turnbull.
\newblock Implementation of {Strassen}'s algorithm for matrix multiplication.
\newblock In {ACM}, editor, {\em Supercomputing '96 Conference Proceedings:
  November 17--22, Pittsburgh, {PA}}. ACM Press and IEEE Computer Society
  Press, 1996.
\newblock \url{www.supercomp.org/sc96/proceedings/SC96PROC/JACOBSON/}.

\bibitem{Huss-Lederman:1996:mai}
S.~Huss-Lederman, E.~M. Jacobson, J.~R. Johnson, A.~Tsao, and T.~Turnbull.
\newblock {Strassen}'s algorithm for matrix multiplication~: Modeling analysis,
  and implementation.
\newblock Technical report, Center for Computing Sciences, Nov. 1996.
\newblock CCS-TR-96-17.

\bibitem{Ibarra:1982:LSP}
O.~H. Ibarra, S.~Moran, and R.~Hui.
\newblock A generalization of the fast {LUP} matrix decomposition algorithm and
  applications.
\newblock {\em Journal of Algorithms}, 3(1):45--56, Mar. 1982.

\bibitem{JeannerodPernet:2007:LQUP}
C.-P. Jeannerod, C.~Pernet, and A.~Storjohann.
\newblock Fast {Gaussian} elimination and the {PLUQ} decomposition.
\newblock Technical report, 2007.

\bibitem{Kreczmar:1976:Strassen}
A.~Kreczmar.
\newblock On memory requirements of {Strassen}'s algorithms.
\newblock In A.~Mazurkiewicz, editor, {\em Proceedings of the 5th Symposium on
  Mathematical Foundations of Computer Science}, volume~45 of {\em LNCS}, pages
  404--407, Gda{\'n}sk, Poland, Sept. 1976. Springer.

\bibitem{Laderman:1992:PAA}
J.~Laderman, V.~Pan, and X.-H. Sha.
\newblock On practical algorithms for accelerated matrix multiplication.
\newblock {\em Linear Algebra and its Applications}, 162--164:557--588, 1992.

\bibitem{Pernet:2001:Winograd}
C.~Pernet.
\newblock Implementation of {Winograd}'s fast matrix multiplication over finite
  fields using {ATLAS} level 3 {BLAS}.
\newblock Technical report, Laboratoire Informatique et Distribution, July
  2001.
\newblock {\small \url{ljk.imag.fr/membres/Jean-Guillaume.Dumas/FFLAS}}

\bibitem{Strassen:1969:GENO}
V.~Strassen.
\newblock {Gaussian} elimination is not optimal.
\newblock {\em Numerische Mathematik}, 13:354--356, 1969.

\bibitem{Winograd:1971:tbtmm}
S.~Winograd.
\newblock On multiplication of 2x2 matrices.
\newblock {\em Linear Algebra and Application}, 4:381--388, 1971.

\end{thebibliography}
\newcommand{\SortNoop}[1]{}

\end{document}